\documentclass[prl,twocolumn,showpacs,amsmath,amssymb]{revtex4}

\usepackage{amssymb,graphicx}
\usepackage{dsfont}
\usepackage{subfigure}

\begin{document}
\title{How to decompose arbitrary continuous-variable quantum operations}
\author{Seckin Sefi}
\email{seckin.sefi@mpl.mpg.de}
\author{Peter van Loock}
\email{peter.vanloock@mpl.mpg.de}
\affiliation{Optical Quantum Information Theory Group, Max Planck Institute for the Science of Light, G\"unther-Scharowsky-Str.1/Bau 26, 91058 Erlangen, Germany}
\affiliation{Institute of Theoretical Physics I, Universit\"at Erlangen-N\"urnberg, Staudstr.7/B2, 91058 Erlangen, Germany}

\begin{abstract}
We present a general, systematic, and efficient method for decomposing any given exponential operator of bosonic mode operators,
describing an arbitrary multi-mode Hamiltonian evolution, into a set of universal unitary gates.
Although our approach is mainly oriented towards continuous-variable quantum computation, it may be
used more generally whenever quantum states are to be transformed deterministically, e.g. in quantum control, discrete-variable quantum computation, or Hamiltonian simulation. We illustrate our scheme by presenting decompositions for various nonlinear Hamiltonians including quartic Kerr interactions. Finally, we conclude with two potential experiments utilizing offline-prepared optical cubic states and homodyne detections, in which
quantum information is processed optically or in an atomic memory using quadratic light-atom interactions.
\end{abstract}


\maketitle

\emph{Introduction}-- Since the proposal of quantum computation as a generalization of computer science, an important theoretical challenge has been how to decompose an arbitrary gate into a universal set. The corresponding theory of discrete-variable decompositions is very extensive and mostly employs matrix representations of logic gates utilizing matrix decomposition techniques \cite{Nielsen}. In contrast to discrete-variable theory, there is not an established method to decompose an arbitrary operator in the continuous-variable (CV) regime except the proof-of-principle results on universal gate sets in Refs.~\cite{Lloyd1999,PhysRevLett.88.097904}. In particular,
Ref.~\cite{Lloyd1999} makes use of an exponential operator approximation and proves that by employing certain elementary gate sets (discussed below) one can derive any operator up to a certain error. However, none of these works intend to present a constructive and efficient decomposition recipe.


The problem of decomposition is intrinsically related to the concept of universality. Universality means to have a set of operators that allows you to simulate any operator on a certain Hilbert space through concatenations of the elements of the universal set. For our purpose, achieving universality is then equivalent to decomposing, at least approximately, an arbitrary unitary exponential operator to a set of elementary unitary exponential operators:
\begin{equation}\nonumber
 e^{itH(a,a^{\dag})}=\{e^{it_1H_1(a,a^{\dag})},e^{it_2H_2(a,a^{\dag})},...,e^{it_NH_N(a,a^{\dag})}\}.
\end{equation}
Here, $a$ and $a^{\dag}$ are annihilation and creation operators, respectively, and $\{H_n\}$ are fixed Hermitian functions of mode operators. The coefficients $t_1,t_2,...$ are interaction times of the Hamiltonians and are functions of $t$. Thus, different concatenations of elements of this set for varying interaction times should enable one to simulate an arbitrary operator. We assume that we have access to arbitrary interaction times for the initial set.

In our setting, there
are two important criteria for CV gate decompositions: how systematic and how efficient the decompositions are. Here we shall derive methods according to these criteria and present a systematic and efficient framework for decomposing any given unitary operator that acts on bosonic modes into a universal set of elementary CV gates. Our general method consists of first expressing operators in terms of linear combinations of commutation operators and then realizing each commutation operator and their combinations through approximations. We will discuss the efficiency of the decompositions and present guidelines to obtain an arbitrary order of error. For this purpose, we employ a powerful technique for obtaining efficient approximations (see supplemental material for details).
Throughout, we use the convention $\hbar=1/2$, i.e., the fundamental commutation relation is $[X,P]=i/2$ with $X\equiv (a^\dag+a)/2$ and $P\equiv i(a^\dag-a)/2$.

\emph{General Gaussian decompositions}-- For Gaussian operators, i.e., second-order operators, exact and finite decompositions to elementary sets are known (here, order is defined as the polynomial order of the mode operators in the Hamiltonian of a given operator). For example, the Bloch-Messiah decomposition allows to decompose any second-order operator, i.e., any unitary Gaussian operation, to passive linear multi-mode optics, single-mode squeezing, and displacement operations \cite{Braunstein2005}.

In Ref.~\cite{PhysRevLett.88.097904} the following set is presented as a single-mode Gaussian universal set: $\{e^{i\frac{\pi}{2}(X^2+P^2)},e^{it_1X},e^{it_2X^2}\}$, and in Ref.~\cite{ukai2010universal}, similar to the Bloch-Messiah decomposition, a recipe is given to decompose any single-mode transformation of second order to this set with no more than four steps.
These exact decompositions emerge from the fact that mode transformations through second-order unitary operations are linear, and thus, one can utilize matrix representations and matrix decomposition techniques. In fact, for Gaussian operator decompositions, one can find infinitely many elementary sets and decompositions, and instead of those sets above, one may choose the one that suits best the given situation and purpose.

\emph{Universal decompositions}-- In order to decompose an arbitrary single-mode operator in CV systems, it has been shown that, in principle, adding a nonlinear element (of order three or more) to the toolbox is sufficient \cite{Lloyd1999}.
In the present work we use the following set:
\begin{equation}\label{eq:cluster_gates}
\{e^{i\frac{\pi}{2}(X^2+P^2)},e^{it_1X},e^{it_2X^2},e^{it_3X^3}\}.
\end{equation}
This set is not unique, and one may use different Gaussian elements as explained above and a different nonlinear element.
However, this particular set turns out to be useful for describing CV quantum computation
in the one-way model using CV cluster states \cite{Raussendorf,Menicucci}.
Note that one can simplify this elementary set further by omitting the second-order Hamiltonian, since $e^{it^2X^2}=e^{i\frac{2t^4}{27}}e^{i\frac{2t}{3}P}e^{itX^3}e^{-i\frac{2t}{3}P}e^{-itX^3}e^{i\frac{t^3}{3}X}$, and using the Fourier transformation whose action is given in Eq.~\eqref{eq:Fourier_transform}. Even though this simplification has value from an academical point of view, as it reduces the minimal number of elementary gates, we are basically motivated by decomposing an arbitrary gate to a set of experimentally accessible gates.
All second-order gates are relatively easy to implement and replacing them by third-order Hamiltonians will increase the complexity
of the gate sequence. Therefore we shall use the overcomplete set \eqref{eq:cluster_gates} in our decompositions without loss of generality. One may also prefer a further extended set depending on a certain experimental situation in order to reduce the complexity of the decompositions.

In addition, one can obtain {\it some} nonlinear operations through unitary conjugation: $Ue^{itH(a,a^\dag)}U^\dag \rightarrow e^{itH(UaU^\dag,(UaU^\dag)^\dag)}$. An important unitary conjugation is the Fourier transform:
\begin{equation}\label{eq:Fourier_transform}
e^{i\frac{\pi}{2}(X^2+P^2)}e^{itX^m}e^{-i\frac{\pi}{2}(X^2+P^2)}=e^{itP^m}.
\end{equation}
Employing unitary conjugation, with the set \eqref{eq:cluster_gates}, one can now generate certain nonlinear gates exactly. For example, $e^{itX^3}e^{itP^2}e^{-itX^3}=e^{it(P-t\frac{3}{2}X^2)^2}$, which is a fourth-order operator. However, there is only a limited number of such decomposable nonlinear operators, and therefore, for generality, we will make use of the idea of operator approximations.

Besides the abstract notion of universality \cite{Lloyd1999} how can a given unitary exponential operator be decomposed to the elementary set \eqref{eq:cluster_gates} in a systematic and efficient way? Let us first introduce the
available toolbox employed to achieve a decomposition as efficient as possible (while efficiency will be defined and discussed later).
The tools we use for CV gate decompositions include Gaussian operator decompositions, unitary conjugation, and exponential operator relations as well as approximations.
Before proceeding to the general case, let us demonstrate how to realize a particular nonlinear exponential operator using the above tools and the set \eqref{eq:cluster_gates}. A very important example is the Kerr interaction operator. It allows to convert a coherent state into a cat state \cite{Yurke1986} and to realize a controlled quantum gate for qubits \cite{PhysRevA.52.3489}. The one-mode self-Kerr interaction, up to a Gaussian element, corresponds to
$e^{it(X^2+P^2)^2}=e^{it(X^4+X^2P^2+P^2X^2+P^4)}$.
In order to decompose the Kerr operation to the set \eqref{eq:cluster_gates}, we would first write the full Kerr Hamiltonian as a linear combination of commutators and then realize each through operator approximations.
The following relations, together with the Fourier transform \eqref{eq:Fourier_transform}, are enough to realize this gate up to a phase:
$X^4=-\frac{2}{9}[X^3,[X^3,P^2]]$,
$X^2P^2+P^2X^2=-\frac{4i}{9}[X^3,P^3]$.
Thus, it is sufficient to realize the above commutators and their linear combinations (see supplemental material for more details).

Let us now present the general scheme for an arbitrary Hamiltonian.
Obviously, any single-mode Hamiltonian, as a polynomial of bosonic mode operators, consists of operators of the form $cX^mP^n+c^\ast P^nX^m$. We can show that any such operator can be written as a linear combination of commutation operators. First note that $cX^mP^n+c^\ast P^nX^m={\rm Re}(c)(X^mP^n+P^nX^m)+i{\rm Im}(c)[X^m,P^n]$, and then one can derive the following two identities (see supplemental material for a derivation),
\begin{align}
X^m=&-\frac{2}{3(m-1)}[X^{m-1},[X^3,P^2]],\label{eq:x_order_increment}\\
X^mP^n+P^nX^m=&-\frac{4i}{(n+1)(m+1)}[X^{m+1},P^{n+1}]\label{eq:fundm_hamilton}\\
&-\frac{1}{n+1}\sum_{k=1}^{n-1}[P^{n-k},[X^m,P^k]].\nonumber
\end{align}
Equation~\eqref{eq:x_order_increment} is necessary to obtain arbitrary powers of $X$ and $P$ operators with the Fourier conjugation \eqref{eq:Fourier_transform}, and Eq.~\eqref{eq:fundm_hamilton} basically prescribes how to systematically decompose an elementary Hamiltonian to commutation operations of orders of $X$ and $P$ and their combinations where we can use the tools we have.

For multi-mode operators one needs an extended elementary set including an entangling operation \cite{Lloyd1999}, for example  in the optical context it can be the beam-splitter operation. Using Gaussian decomposition methods, for
simplicity, we may assume that we have access to the following gate without loss of generality, $C_Z=e^{2itX_1\otimes X_2}$. For multi-mode Hamiltonians we can again use the simplifications for a single mode and Eq.~\eqref{eq:fundm_hamilton}, because of the fact that the operators on one mode commute with the operators on the other mode. However, for this purpose, we initially need to realize the two-mode operations with arbitrary powers of $X$ and $P$ in both modes, similar to the single-mode relation \eqref{eq:x_order_increment}. The following relation together with Fourier conjugation \eqref{eq:Fourier_transform} and Eq.~\eqref{eq:x_order_increment}, is sufficient to realize the two-mode operations with arbitrary powers of $X$ and $P$,
\begin{equation}\label{eq:multim_fund_hamilton1}
 P_1^{n}\otimes P_2^{s}=-\frac{1}{(n+1)(s+1)}[P_2^{s+1},[P_1^{n+1},X_1\otimes X_2]].
\end{equation}
Then, we can use Eq.~\eqref{eq:fundm_hamilton} again with the single-mode operations to realize an arbitrary two-mode expression.
As an example, consider the cross-Kerr Hamiltonian (up to a Gaussian transformation): $(X^2+P^2)_1\otimes (X^2+P^2)_2=X_1^2\otimes X_2^2+X_1^2\otimes P_2^2+P_1^2\otimes X_2^2+P_1^2\otimes P_2^2$. In this case, the nested commutator $[P_2^3,[P_1^3,X_1\otimes X_2]]$ will suffice.
Another, even simpler example for decomposing a nonlinear two-mode evolution, namely cubic parametric down conversion with a quantized pump, is discussed in the supplemental material. In form of a dispersive interaction, it may also be used to mimic a Kerr gate \cite{Klimov}.

From an academic perspective, Eqs.~\eqref{eq:x_order_increment},\eqref{eq:fundm_hamilton}, and \eqref{eq:multim_fund_hamilton1} are universal not only for any Hamiltonian, but also for any initial universal set with a nonlinear gate different from $X^3$ because of the well-known equations: $\frac{\partial F}{\partial P}=-2i[X,F]$, $\frac{\partial F}{\partial X}=2i[P,F]$, where $F$ is a function of operators $X$ and $P$. Thus, any initial nonlinear Hamiltonian can be reduced to a form $X^m$. From a practical perspective, however, it is unwise to use
Eqs.~\eqref{eq:x_order_increment} and \eqref{eq:fundm_hamilton} for arbitrary initial sets because of a typical increase of complexity. Instead, one should derive an optimized expression (in terms of the decomposition efficiency, see below) utilizing the available tools for every other Hamiltonian and every other universal set.

\emph{Efficiency}-- Besides having a systematic method, we also require the decompositions to be relatively efficient. We define efficiency as the number of operators needed to realize, in an approximate fashion, a given operator with a certain negligible error (note that this definition slightly differs from
previous ones \cite{Lloyd1999} where efficiency is the scaling of the number of operators with respect to the error).
Let us give a few more definitions. If in
\begin{equation}\label{eq:gbch_problem}
e^{tC}=e^{t_1A}e^{t_2B}e^{t_3A}...e^{t_M B},
\end{equation}
the Taylor expansion of both sides matches for the orders of $t$ up to $t^m$, it is called $m$th-order decomposition \footnote{The approximation quality does not only depend on the decomposition {\it order}, but also on the actual {\it values} of the gate interaction times and the {\it norms} of the corresponding operators. In the present work, the focus is on the decomposition order, omitting the absolute errors.}.
For example, an important case is when $C=A+B$ for which we will use the term splitting. Another important case is when $C=[A,B]$ which, from now on, we call commutation operator. For example, the identity below is a well-known second-order approximation for a commutation operator. It has been used already in quantum control \cite{clark2003control}, discrete-variable quantum computation \cite{Lloyd1995}, CV quantum computation \cite{Lloyd1999}, or, in general, Hamiltonian simulation theory \cite{brown2010using},
\begin{equation}
e^{t^2[A,B]}=e^{itB}e^{itA}e^{-itB}e^{-itA}+f(t^3,A,B)+\ldots\label{eq:bch3}\\
\end{equation}
It basically says that, for $t<1$, the corresponding operator concatenation is the same as applying the commutation operator of $A$ and $B$, up to some error where the dominant term is of the order $t^3$.
Now in order to obtain more reliable gates, a straightforward and common way to improve accuracy is using smaller interaction times, $t\rightarrow t/n$, and applying the decomposition $n^2$ times to obtain the same interaction time as before,
\begin{equation}\label{eq:sec_ord_impro_approx}
 e^{t^2[A,B]}=\left(e^{iB\frac{t}{n}}e^{iA\frac{t}{n}}e^{-iB\frac{t}{n}}e^{-iA\frac{t}{n}}\right)^{n^2}+f\left(\frac{t^3}{n},A,B\right).\nonumber
\end{equation}
Let us call this approach rescaling (\cite{suzuki1990fractal} and Refs. therein). Besides improving accuracy, rescaling is absolutely necessary to realize nested commutations. For example, one may replace $itA$ by $t^2[B,A]$ in Eq.~\eqref{eq:bch3} to simulate the nested commutation operator $e^{it^3[B,[B,A]]}=e^{itB}e^{t^2[B,A]}e^{-itB}e^{-t^2[B,A]}+f'(t^4,A,B)$, and similarly for further nested commutations. However, in this identity, an approximation of $[B,A]$ is still needed and eventually, using again Eq.~\eqref{eq:bch3}, we obtain an operator whose interaction time is of the same order as the dominant error term. In order to obtain a reasonable decomposition, the order of the dominant error should be smaller than $t^3$. Thus, again rescaling is needed, requiring relatively many operators to enhance accuracy.
For instance, using the approximations described so far, the number of operators to approximate a single commutation operator with coefficient 0.1 and dominant error term $\sim 10^{-3}$ requires 4000 operators, while for the nested commutation operator, with the same values, the number of operators will be $\sim 10^{10}$ (see supplemental material). Hence, better approximations are crucial to achieve reasonable decompositions.

What we propose to use is a general and powerful method for obtaining higher-order approximations in a direct fashion (details on this rather technical part of our proposal can be found in the supplemental material; see also, for instance, page 393 of Ref.~\cite{McLachlan2002}). In this approach, concatenations converge much faster to an arbitrary set of commutations and linear combinations of these, reducing the number of operators from the order of $10^{10}$ to the order of tens.

\emph{Experiment}-- An immediate consequence of our work is that it bridges the originally huge gap between abstract notions of CV decomposition theory and possible experimental implementations, for instance, within quantum optics. Highly nonlinear quantum gates (such as quartic Kerr gates) may then be realized in a deterministic fashion by concatenating tens of quadratic and cubic gates. Figure~1 illustrates two experimental schemes.
They both combine ideas for conditional optical state preparation, in order to probabilistically prepare and distill high-fidelity cubic phase states
$\approx \int dx e^{it x^3} |x\rangle$ \cite{GKP,Ghose,Marek}, with those for temporally encoded \cite{MenicucciRalph,Acin} and fully homodyne-detection-based \cite{Gu} CV cluster computation.
The complication of realizing nonlinear gates is shifted offline into the preparation of the cubic ancillae.

\begin{figure}[t]
\centering
{
\includegraphics[width=7.5cm, height=5cm]{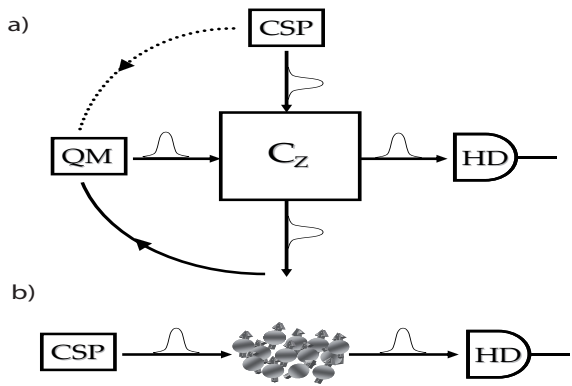}
}
\caption{
Concatenating elementary gates using offline conditional state preparation (CSP)
of squeezed and cubic states, a quantum memory (QM), an optical $C_Z$ gate, $e^{2itX_1\otimes X_2}$,
and homodyne detection (HD); solid lines represent the optical paths.
In a), an optical quantum state is released from the QM
only when the CSP succeeds (dotted line is classical feedforward). In b), quantum information storage and
processing go hand in hand by employing suitable quadratic light-matter interactions between optical pulses
and an atomic ensemble.
}
\label{fig}
\end{figure}

In Fig.~1a), the conditional state preparer (CSP) is linked with a quantum memory (QM) through a classical channel in order to signal whenever an offline state is available such that an optical
pulse carrying the latest quantum information is released from the quantum memory. After the first application of the optical $C_Z$ gate between a first input pulse coming from the left and a first ancilla pulse coming from the top (emerging from the CSP), the $C_Z$-transformed input pulse
is measured through homodyne detection (HD), while the $C_Z$-transformed ancilla pulse is sent to the quantum memory with its quantum state transferred
onto the memory. After preparation of the next ancilla pulse, a new input pulse emerges from the memory for a second application of the
$C_Z$ gate and so on. This is similar to the scheme of Ref.~\cite{MenicucciRalph} for a single quantum wire (corresponding to a linear CV cluster chain), but this time including cubic states and quantum memories, and excluding non-linear measurements such as photon counting (except for the CSP).
Note that storage of non-classical states has been already demonstrated experimentally \cite{Eisaman,Chaneliere}.

The scheme in Fig.~1b) uses an atomic ensemble and a quadratic light-matter interaction \cite{Polzik}.
The light-matter interaction is always delayed until an appropriate offline state is available.
Quantum information can be stored and processed at the same time within the
ensemble \cite{Acin} by swapping the optical and atomic states after every interaction, simply using additional quadratic interactions
\cite{Fiurasek,Fleischhauer}.
As opposed to Refs.~\cite{MenicucciRalph,Acin}, the optical state is either a squeezed vacuum or a conditionally prepared cubic state, inserted into the temporal cluster chain whenever needed \cite{Gu}. Thus, we make explicit use of the atomic memory
to compensate for the probabilistic optical ancilla-state preparations. Only with these cubic ancillae is it possible to perform universal gates solely through homodyne detection \cite{Gu}. In principle, our decomposition method would then determine what homodyne local-oscillator angles to choose
in order to realize a given Hamiltonian in a deterministic fashion.
Currently available squeezing levels
of almost 13dB are, in principle, sufficient to perform up to 73 elementary teleportations in a nonclassical fashion; four such elementary Gaussian gates have been implemented already in a fully optical, spatial encoding using 5.5dB-squeezed light sources \cite{UkaiPRL}.

In summary, we proposed a systematic framework to decompose arbitrary CV unitaries (from arbitrary multi-mode Hamiltonians) into an experimentally realizable elementary gate set.
Different from previous proof-of-principle demonstrations, our treatment brings the abstract notions of decomposition theory for CV quantum computation close to experimental implementations.




We acknowledge support from the Emmy Noether Program of the DFG in Germany and thank Nick Menicucci and Akira Furusawa
for useful discussions.

\onecolumngrid
\section{SUPPLEMENTARY MATERIAL}
\section{Derivation of Equation (4)}

\begin{align*}
[X^m,P^n]&=X^mP^n-P^nX^m\\
&=X^mP^n-P^{n-1}(X^mP-\frac{im}{2}X^{m-1})\\
&=X^mP^n+\frac{im}{2}P^{n-1}X^{m-1}-P^{n-2}PX^mP\\
&=X^mP^n+\frac{im}{2}P^{n-1}X^{m-1}-P^{n-2}X^mP^2+\frac{im}{2}P^{n-2}X^{m-1}P\\
&=X^mP^n+\frac{im}{2}P^{n-1}X^{m-1}+\frac{im}{2}P^{n-2}X^{m-1}P-P^{n-2}X^mP^2\\
\vdots\\
&=\frac{im}{2}\sum_{k=0}^{n-1} P^kX^{m-1}P^{n-k-1}
\end{align*}

\begin{align*}
[X^m,P^n]&=\frac{im}{4}\sum_{k=0}^{n-1} P^kX^{m-1}P^{n-k-1}+P^{n-k-1}X^{m-1}P^k\\
&=\frac{im}{4}\sum_{k=0}^{n-1} \left( (X^{m-1}P^k-[X^{m-1},P^k])P^{n-k-1} + P^{n-k-1}([X^{m-1},P^k]+P^kX^{m-1}) \right)\\
&=\frac{im}{4}\sum_{k=0}^{n-1} \left( X^{m-1}P^{n-1}-[X^{m-1},P^k]P^{n-k-1} + P^{n-k-1}[X^{m-1},P^k]+P^{n-1}X^{m-1} \right)\\
&=\frac{im}{4}\sum_{k=0}^{n-1} \left( X^{m-1}P^{n-1}+P^{n-1}X^{m-1}+ [P^{n-k-1},[X^{m-1},P^k]] \right)\\
&=\frac{imn}{4} \left( X^{m-1}P^{n-1}+P^{n-1}X^{m-1}\right)+ \frac{im}{4}\left( \sum_{k=1}^{n-2} [P^{n-k-1},[X^{m-1},P^k]] \right)\\
\end{align*}
As a result:
\begin{equation*}
X^mP^n+P^nX^m=-\frac{4i}{(n+1)(m+1)}[X^{m+1},P^{n+1}] -\frac{1}{n+1}\sum_{k=1}^{n-1}[P^{n-k},[X^m,P^k]]
\end{equation*}
Note that for $n=1$, the summation term in the identity above is zero. Also, due to the Jacobi identity, we have $[P^{n-k},[X^m,P^k]]=[P^{k},[X^m,P^{n-k}]]$, and this may also lead to some simplification depending on the value of $n$.

\section{Necessity of good approximations}

Here, we shall illustrate the necessity for having better (than any commonly used) approximations.
The approximations that we employ are explicitly introduced below in this supplemental material.

For instance, for a nested commutation approximation of an interaction strength $0.1$ and a dominant error term of $10^{-3}$, we would need $73$ operations corresponding to an eighth-order approximation. For comparison, note that it is also possible to use Lloyd's method \cite{Lloyd1995}. Lloyd's idea was originally intended as a proof of principle, but it has also been used in the literature as an approximation tool. The required number of operators using Lloyd's method is evaluated below. The notation $N[x]$ is used to indicate that a number of $x$ operators is required.
For the commutation operator, we have
\begin{align}
e^{t^2[A,B]}&=\left(e^{iB\frac{t}{n}}e^{iA\frac{t}{n}}e^{-iB\frac{t}{n}}e^{-iA\frac{t}{n}}\right)^{n^2}+f\left(\frac{t^3}{n},A,B\right)\label{basic_comm}\\
e^{t[A,B]}&=N[4\times n^2]+f\left(\frac{t^{3/2}}{n}\right)\nonumber
\end{align}
while for the nested commutation, we obtain
\begin{align}
e^{it^3[B,[B,A]]}&=e^{itB}e^{t^2[B,A]}e^{-itB}e^{-t^2[B,A]}+f'(t^4,A,B)\label{basic_nested}\\
&=\left(e^{i\frac{t}{m}B}e^{\frac{t^2}{m^2}[B,A]}e^{-i\frac{t}{m}B}e^{-\frac{t^2}{m^2}[B,A]}\right)^{m^3}+f'\left(\frac{t^4}{m}\right)\nonumber\\
&=\left(e^{i\frac{t}{m}B}\left(e^{iB\frac{t}{ml}}e^{iA\frac{t}{ml}}e^{-iB\frac{t}{ml}}e^{-iA\frac{t}{ml}}\right)^{l^2}e^{-i\frac{t}{m}B}
\left(e^{-iB\frac{t}{ml}}e^{-iA\frac{t}{ml}}e^{iB\frac{t}{ml}}e^{iA\frac{t}{ml}}\right)^{l^2}\right)^{m^3}+
f'\left(\frac{t^4}{m}\right)+f\left(\frac{t^3}{l}\right)\nonumber\\
e^{it[B,[B,A]]}&=N[8\times l^2\times m^3]+f'\left(\frac{t^{4/3}}{m}\right)+f\left(\frac{t}{l}\right)\nonumber
\end{align}
Thus, using these approximations, the number of (elementary) operations will be of the order of $10^{10}$
for an interaction strength of $0.1$ and a dominant error term of $10^{-3}$.

\section{Example: one-mode self-Kerr interaction gate}

For the decomposition of a Kerr gate with coefficient 0.1 and dominant error term $10^{-3}$, we need to approximate the following operators: $O_1=-\frac{2}{9}[X^3,[X^3,P^2]]$, $O_3=-\frac{4i}{9}[X^3,P^3]$.
The Kerr gate is then : $e^{i0.1(O_1+O_2+O_3)}$. Here $O_2$ is the Fourier transform of $O_1$, i.e., $e^{itO_2}=Fe^{itO_1}F^\dag$. First, using second-order three-party splitting (see
Refs.~\cite{suzuki1990fractal,yoshida1990construction}), we split into separate elements:
\begin{equation}
e^{i0.1(O_1+O_2+O_3)}=e^{i0.05O_1}e^{i0.05O_2}e^{i0.1O_3}e^{i0.05O_2}e^{i0.05O_1} + 10^{-3}\times f\left(O_1,O_2,O_3\right)+\ldots\label{sequence}
\end{equation}
and then we insert the approximations for the commutation and the nested commutation operators (see below). Approximations for these operators can be calculated considering the necessary approximation order. For the nested commutation operator with coefficient $0.05\times 2/9$ and dominant error term smaller than $10^{-3}$, we need the fourth-order approximation which corresponds to 9 operators,
\begin{eqnarray*}
e^{-i0.05\times \frac{2}{9}[X^3,[X^3,P^2]]}&=&e^{-i0.11157P^2}e^{i0.02231X^3}e^{i0.02231P^2}e^{-i0.02231X^3}e^{-i0.02231P^2}\\&&e^{-i0.02231X^3}e^{i0.02231P^2}e^{i0.02231X^3}e^{-i0.11157P^2} + 0.55326\times 10^{-3}\times f(X^3,P^2) + ...
\end{eqnarray*}
Similarly,
the commutation operator with coefficient $0.1\times 4/9$ and dominant error term smaller than $10^{-3}$ also requires the fourth-order approximation, this time corresponding to 10 operators,
\begin{eqnarray*}
e^{-0.1\times \frac{4}{9}[X^3,P^3]}&=&e^{-i0.25298X^3}e^{i0.210819P^3}e^{i0.01918X^3}e^{-i0.28476P^3}e^{i0.36163X^3}\\&&e^{i0.36053P^3}e^{i0.12861X^3}e^{-i0.05805P^3}e^{-i0.25644X^3}e^{-i0.22853P^3} + 0.41643\times 10^{-3}\times f(X^3,P^3) + ...
\end{eqnarray*}
As we have to apply $O_1$ four times and $O_3$ only once, we require $4\times 9 + 10 = 46$ elementary operations.
Every single $O_1$ operation as well as the $O_3$ operation each consume $\sim 10$ additional Fourier transforms in order to switch from the elementary $X$ gates to the necessary $P$ gates [here, $\sim 10$ indicates that some Fourier gates cancel in the sequence \eqref{sequence} when switching between $O_1$ and $O_2$]. As a result, summing up, we will need $9+10+9+11+9 = 48$ extra Fourier gates,
in addition to the $46$ elementary quadratic and cubic $X$ gates. In total, this leads to $94$ elementary operations from the universal
gate set needed to simulate a Kerr interaction of size $0.1$ with errors scaling smaller than $10^{-3}$.
Without using these more powerful approximation techniques (as explained in detail below) and using instead standard techniques such as the well-known Trotter formula for splitting and the standard approximations for the commutation operator \eqref{basic_comm} and the nested commutation operator \eqref{basic_nested}, we would need $\sim 10^8$ operations
to simulate this Kerr interaction gate with the same precision.

\section{Example: parametric-down-conversion Hamiltonian with quantized pump}

As a second example, we discuss the decomposition of the Hamiltonian evolution of parametric down conversion with a quantized pump.
This example illustrates that one can sometimes find exact operator relations and thus reduce the use of operator approximations to some extent.
The operator to be decomposed is: $e^{it(X_1^2X_2-P_1^2X_2+X_1P_1P_2+P_1X_1P_2)}$, where the subscripts denote the two optical modes. As described, we first divide the Hamiltonian into its simplest Hermitian parts $\{e^{iX_1^2X_2},e^{iP_1^2X_2},e^{i(X_1P_1P_2+P_1X_1P_2)}\}$ through splitting approximations. Now the operators $e^{iX_1^2X_2}$ and $e^{iP_1^2X_2}$ can be realized through exact operator relations instead of operator approximations:
\begin{equation}
e^{-ikP_2^3}e^{i\alpha X_1X_2}e^{ikP_2^3}e^{-2i\alpha X_1X_2}e^{ikP_2^3}e^{i\alpha X_1X_2}e^{-ikP_2^3}=e^{\frac{3}{2}ik\alpha ^2X_1^2P_2}
\end{equation}
Here, $k$ and $\alpha$ are some free parameters and $e^{iX_1^2X_2},e^{iP_1^2X_2}$ are equivalent to $e^{iX_1^2P_2}$ up to Fourier transforms. The only remaining operator, $e^{i(X_1P_1P_2+P_1X_1P_2)}$, can be realized through a commutation approximation of $[P_1^2,X_1^2P_2]$.

\section{Approximating commutation and nested commutation operators}

\subsection{Introduction}

In this section, we shall explain a very powerful method that we propose to employ for deriving the necessary approximations in our scheme. We also present explicitly some efficient approximations which are used throughout the paper.

Let us first define the problem. Suppose we have access to a finite set of non-commuting exponential operators, $\{e^{t'\hat{A}},e^{t''\hat{B}},e^{t'''\hat{C}},...\}$,
where $\hat A$, $\hat B$, $\hat C$,... are arbitrary operators.
Through different and finite concatenations of these operators, our aim is to obtain another operator, $\{e^{t\hat{O}}\}$, approximately, i.e., up to a certain error,
\begin{equation}\label{eq:problem}
e^{t\hat{O}}=\prod_i^M e^{t_i\hat{A}}e^{t_i'\hat{B}}e^{t_i''\hat{C}}...
\end{equation}
where $t,t_i,t_i',...$ are scalars. Also, we further assume that $t_i,t_i',...$ are not fixed, i.e., they are adjustable parameters being functions of $t$ as a result of the approximation problem. If the Taylor expansion of both sides of Eq.~\eqref{eq:problem} matches for the orders of $t$ up to $t^m$, then the right hand side of the equation is called $m$'th order approximation for the operator $e^{t\hat{O}}$, and the problem is to find a method for this approximation, i.e., to find the appropriate values for $t_i,t_i',...$.

In the present context, we are interested in the case where $\hat{O}$ is an arbitrary nested commutation operator comprised of $\hat{A}$ and $\hat{B}$: $[.,[.,[.,...[\hat{A},\hat{B}]]]]$, which is to be approximated through the set $\{e^{t\hat{A}},e^{t'\hat{B}}\}$. Throughout this section we call an exponential operator with exponent $[\hat{A},\hat{B}]$ \emph{commutation operator}, and similarly for the nested commutators. In general, we omit hats to indicate operators except when confusions are possible.

\subsection{Employed method}

The method we employ combines various elements of existing splitting and approximation techniques (see, for instance, page 393 of the splitting review Ref.~\cite{McLachlan2002}). Our execution of these methods relies upon the generalized Baker-Campbell-Hausdorff series (gBCH) which is basically the logarithm of an arbitrary exponential operator concatenation involving more than two operators; an extension of the standard BCH, as the name indicates.
It is possible to calculate the gBCH in various ways, for instance, by repeatedly utilizing formulas for the standard BCH. With the aid of computer software, this is straightforward. In the present work, however, we shall employ a very convenient calculation technique proposed by Reinsch \cite{reinsch2000simple}.

For any necessary orders of the operator approximation, we use two calculation steps. In the first step, we employ a brute force and hence efficient method for approximations. In the second step, we obtain higher-order approximations using the results of the first step. This second step will be necessary in order to be able to achieve orders of approximations otherwise unattainable through the
first step alone.

In order to derive a certain order of approximation, we start with a specific operator concatenation like the one in Eq.~\eqref{eq:problem}. We calculate the logarithm of this concatenation through the gBCH and this results in a linear combination of a set of operators. The coefficients of these operators correspond to polynomials of the input parameters. We then solve all the polynomial equations in order to eliminate the undesired operators in the linear combination. The solutions of the polynomial equation set give us the proper values of the input parameters for a particular approximation. Hence, the whole problem of obtaining a specific order of approximation is reduced to a computationally well defined problem, namely that of solving polynomial equations.

For example, consider a fourth-order approximation of the commutation operator $e^{t^2[A,B]}$ through the operators $\{e^{it'A},e^{it''B}\}$. Here, $i$ is the square root of -1. We start from a concatenation of alternating $e^{it'A}$ and $e^{it''B}$ operators. The concatenation consists of ten operators, for which the reason will become apparent below. Thus, we have
\begin{eqnarray}\label{eq:com_example}
\begin{split}
\log (e^{c_1itA}e^{c_2itB}e^{c_3itA}...e^{c_{10}it B})=&f_1tA+f_2tB+f_3t^2[A,B]+f_4t^3[A[A,B]]+f_5t^3[B,[A,B]]\\
&+f_6t^4[A,[A,[A,B]]]+f_7t^4[B,[A,[A,B]]]+f_8t^4[B,[B,[A,B]]]+...
\end{split}
\end{eqnarray}
where the linear dependent operators such as $[B,A]$ are omitted. The eight coefficients, $\{f_1,...,f_8\}$, are polynomials of the input parameters, $\{c_1,...,c_{10}\}$. To these parameters, we assign the values $c_1=1.2$ and $c_2=-1$ to ensure a real solution set. Then we solve these polynomials simultaneously with the conditions $f_3=1$ and $f_k=0$, $\forall k\neq 3$. As a result, we obtain sets of appropriate values for the input parameters. Consequently, the left hand side of Eq.~\eqref{eq:com_example} can be considered a fourth-order approximation for the commutation operator. More
specifically, for the first step of the calculation, we need to take into account the following points.

{\it Constructing real solutions--} the number of operators in the concatenation should always match the number of polynomials to be solved. However, the solutions might be all complex, which is useless for physical scenarios. In this case, one has to add more operators to the concatenation to be able to adjust the input parameters and obtain an all real-valued solution set. This is why in the above example, we added two more operators to the concatenation and chose the specific values 1.2 and -1. We choose the number of extra operators and the assigned values by trial and error.

{\it Removing linear dependency--} one should reduce the polynomials for simplification. The polynomial orders depend on the order of the operators that belong to the polynomials (for instance, in the above example, $f_3$ is a second-order polynomial, but $f_4$ and $f_5$ are third-order polynomials), and for every order there exist many linear dependent polynomials. Especially for higher orders, these will consist of thousands of elements. Thus, in order to reduce the complexity, one should eliminate the linear dependent polynomials. For example, a concatenation series comprised of 14 operators, for a fifth-order approximation, in general, gives rise to 51 polynomials, but, after the elimination of the linear dependent ones, this reduces to 14.

{\it Solving polynomial equations--} as an algorithm for polynomial equation solving, we do not recommend to employ a conventional analytical method such as those standard methods based upon Gr{\"o}bner bases or other standard commercial computer software. These approaches hardly help for obtaining solutions of orders higher than four. Instead, for polynomial equation solving, we use the so-called homotopy continuation method \cite{Morgan2009} for the first step of the calculation scheme (as well as for the second step, as described later).

The homotopy continuation method is an iterative numerical method. Even though it has its own complications and subtleties, we shall only use the basic principle of that method. We employ it for iterating Newton's polynomial solving algorithm. Newton's algorithm works only provided one can guess the roots of polynomials very close to their actual values. Homotopy continuation is basically a general idea to circumvent this kind of problems by just starting from a simple toy problem and slowly converting it into the real problem while tracing the solutions. In our case, we start with a set of polynomial equations the roots of which are obvious, and make small changes on these initial polynomials. We then solve the new equations using the results from the previous polynomial solving step as the starting point for Newton's method. Further, we iterate this process such that at some point we get the original polynomials that we wanted to solve, and in addition some of their solutions. Besides the overall time efficiency of this approach, the method allows us to deal with each solution separately, thus, eliminating the need of finding all the solutions. This property indirectly improves the efficiency dramatically.

We can obtain solutions for up to fifth-order approximations and with a little more expertise on polynomial equation solving and patience, it should be possible to obtain even sixth-order approximations. However, in order to obtain higher-order approximations, we believe that the first step
alone is not sufficient.
Therefore, we propose to employ yet another step in order to be able to go to higher orders 

In order to circumvent the problem of having an intractable calculation, we have
to employ another step.
In this second step, to derive higher order approximations, we now use a concatenation of those approximations that we obtain through the first step. This second step is similar to the Suzuki-Yoshida method \cite{suzuki1990fractal,Suzuki1992b,yoshida1990construction}. The difference is, instead of improving step by step and obtaining every higher order recursively, here, by utilizing the gBCH, we obtain a by many orders higher approximation in a single step.

Let us turn again to our example of commutation operators. Suppose we want to obtain a ninth-order approximation for a given commutation operator. In this case, we can obtain the fifth order with the first step, corresponding to a approximation for an operator of the form $e^{t^2[A,B]+t^6\hat{F}+t^{7}\hat{G}+...}$. Let us denote this as $Q_5(t)$. We then take a concatenation series of two such operators $Q_5(t)$ and $Q_5^{-1}(t)$. Obviously, the approximation of $Q_5^{-1}(t)$ is the reverse ordered approximation of $Q_5(-t)$, and hence is an operator of the form $e^{-t^2[A,B]-t^6\hat{F}-t^{7}\hat{G}-...}$. The concatenation we use is $\prod_{i} Q_5(p_it)Q_5^{-1}(p_i't)$ where $p_i$ and $p_i'$ are parameters that we have to find (this kind of concatenation is also used in Ref.~\cite{Suzuki1992b} for a recursive approximation in the splitting problem). We again employ the gBCH to obtain the resultant operator combination and eliminate the operators from order six to nine. For this particular example, we need to solve seven polynomials, but due to the constraint of a real solution set, we use nine operators. When we rewrite each fifth-order operator in terms of their approximated form, in total, we need 144 operators.
Obviously, this method also heavily depends on the efficiency of polynomial solving.

\subsection{Examples}\label{examples}

In this section, we present some commutation and nested commutation approximations. Although for each case there exist many solution sets, we always
present only one of these and we only give the explicit numbers up to six digits out of fifteen digits. Also, note that here some parameters are indicated as zero in order to avoid confusion in the general concatenation ansatz that we used.


\subsubsection{Commutation operator}

Here, the operator to be approximated is $e^{t^2[A,B]}$ and the general ansatz we use is:

\begin{equation}\label{gen_patt}
\prod_i^M e^{c_iitA}e^{c_i'itB}
\end{equation}

A fourth-order and fifth-order approximation is presented in table \ref{tbl:fourth order commutation operator} and table \ref{tbl:fifth order commutation operator}, respectively.

\begin{table}[htb!]
\begin{center}
\begin{tabular}{l|l|l}
$i$ & $c_i$ & $c_i'$ \\
 \hline
1 & 1.2 & -1 \\
2 & -0.090992 & 1.350762 \\
3 & -1.715364 & -1.710162 \\
4 & -0.610065 & 0.275377 \\
5 & 1.216422 & 1.084021
\end{tabular}
\end{center}
\caption{fourth-order commutation operator}
\label{tbl:fourth order commutation operator}
\end{table}

\begin{table}[htb!]
\begin{center}
\begin{tabular}{l|l|l}
$i$ & $c_i$ & $c_i'$ \\
 \hline
1 & 1.2 & -1 \\
2 & -2.121328 & -1.680681 \\
3 & -0.0199879 & 1.602050 \\
4 & 2.025319 & 2.743677 \\
5 & 0.061217 & -1.170812  \\
6 & -3.235890 & 0.018735  \\
7 & 1.519023 & -0.989125  \\
8 & 0.571646 & 0.476156
\end{tabular}
\end{center}
\caption{fifth-order commutation operator}
\label{tbl:fifth order commutation operator}
\end{table}

\begin{table}[htb!]
\begin{center}
\begin{tabular}{l|l|l}
$i$ & $p_i$ & $p_i'$ \\
 \hline
1 & 0.8 & -1 \\
2 & 1.427601 & 1.727306 \\
3 & 1.816745 & 1.700274 \\
4 & 1.371191 & -1.693846 \\
5 & -1.698488 & 0
\end{tabular}
\end{center}
\caption{ninth-order commutation operator approximation through fifth-order approximation}
\label{tbl:ninth order commutation operator through fifth order decomposition}
\end{table}

We can then employ the second step of the calculation scheme to generate a ninth-order approximation through fifth-order approximations, as shown
in table \ref{tbl:ninth order commutation operator through fifth order decomposition}. For this purpose, the concatenation ansatz we use is:

\begin{equation}\label{eq:second_gen_patt}
 \prod_{i} Q_5(p_it)Q_5^{-1}(p_i't)
 \end{equation}

\subsubsection{Nested commutation operator}

Here, the operator to be approximated is $e^{it^3[A,[A,B]]}$ and the general ansatz we use is that of Eq.~\eqref{gen_patt}.
Respectively, in table \ref{tbl:fourth order nested commutation operator} and in table \ref{tbl:fifth order nested commutation operator},
fourth-order and fifth-order approximations are presented.

\begin{table}[htb!]
\begin{center}
\begin{tabular}{l|l|l}
$i$ & $c_i$ & $c_i'$ \\
 \hline
1 & 0 & 0.5 \\
2 & -1 & -1 \\
3 & 1 & 1 \\
4 & 1 & -1 \\
5 & -1 & 0.5
\end{tabular}
\end{center}
\caption{fourth-order nested commutation operator}
\label{tbl:fourth order nested commutation operator}
\end{table}

\begin{table}[htb!]
\begin{center}
\begin{tabular}{l|l|l}
$i$ & $c_i$ & $c_i'$ \\
 \hline
1 & 0 & 0.5 \\
2 & -0.912433 & -1.000141 \\
3 & 2.439891 & -0.531940 \\
4 & 0.184788 & 2.000998 \\
5 & -0.477395 & -1.659152 \\
6 & -0.761744 & 1.465192 \\
7 & 1.181123 & -1.312200 \\
8 & -1.654230 & 0.537205
\end{tabular}
\end{center}
\caption{fifth-order nested commutation operator}
\label{tbl:fifth order nested commutation operator}
\end{table}

In table \ref{tbl:ninth order nested commutation operator through fifth order decomposition}, a ninth-order approximation is presented using the ansatz of Eq.~\eqref{eq:second_gen_patt}.

\begin{table}[htb!]
\begin{center}
\begin{tabular}{l|l|l}
$i$ & $p_i$ & $p_i'$ \\
 \hline
1 & 1 & 1.2 \\
2 & -0.948048 & -0.830428 \\
3 & 1.049237 & -0.887661 \\
4 & 1.358200 & 1.329989
\end{tabular}
\end{center}
\caption{ninth-order nested commutation operator through fifth-order approximation}
\label{tbl:ninth order nested commutation operator through fifth order decomposition}
\end{table}

Given an initial approximation, it is also possible to use the symmetry argument of Suzuki \cite{suzuki1990fractal,Suzuki1992b} and construct recursively symmetrical approximations for the nested commutation operator, since $e^{it^3[A,[A,B]]}e^{i(-t)^3[A,[A,B]]}=I$.
Hence, a sixth-order approximation for nested commutation can be constructed through fourth orders using $3\times 9-2=25$ operations.


\begin{thebibliography}{20}
\expandafter\ifx\csname natexlab\endcsname\relax\def\natexlab#1{#1}\fi
\expandafter\ifx\csname bibnamefont\endcsname\relax
  \def\bibnamefont#1{#1}\fi
\expandafter\ifx\csname bibfnamefont\endcsname\relax
  \def\bibfnamefont#1{#1}\fi
\expandafter\ifx\csname citenamefont\endcsname\relax
  \def\citenamefont#1{#1}\fi
\expandafter\ifx\csname url\endcsname\relax
  \def\url#1{\texttt{#1}}\fi
\expandafter\ifx\csname urlprefix\endcsname\relax\def\urlprefix{URL }\fi
\providecommand{\bibinfo}[2]{#2}
\providecommand{\eprint}[2][]{\url{#2}}

\bibitem{Nielsen} M.\ A.\ Nielsen and I.\ L.\ Chuang,
{\it Quantum Computation and Quantum Information},
Cambridge (2000).

\bibitem[{\citenamefont{Lloyd and Braunstein}(1999)}]{Lloyd1999}
\bibinfo{author}{\bibfnamefont{S.}~\bibnamefont{Lloyd}} \bibnamefont{and}
  \bibinfo{author}{\bibfnamefont{S.}~\bibnamefont{Braunstein}},
  \bibinfo{journal}{Phys. Rev. Lett.} \textbf{\bibinfo{volume}{82}},
  \bibinfo{pages}{1784} (\bibinfo{year}{1999}).

\bibitem[{\citenamefont{Bartlett et~al.}(2002)\citenamefont{Bartlett, Sanders,
  Braunstein, and Nemoto}}]{PhysRevLett.88.097904}
\bibinfo{author}{\bibfnamefont{S.~D.} \bibnamefont{Bartlett {\it et al.}}},
  \bibinfo{journal}{Phys. Rev. Lett.} \textbf{\bibinfo{volume}{88}},
  \bibinfo{pages}{097904} (\bibinfo{year}{2002}).


\bibitem[{\citenamefont{Braunstein}(2005)}]{Braunstein2005}
\bibinfo{author}{\bibfnamefont{S.}~\bibnamefont{Braunstein}},
  \bibinfo{journal}{Phys. Rev. A} \textbf{\bibinfo{volume}{71}},
  \bibinfo{pages}{55801} (\bibinfo{year}{2005}).

\bibitem[{\citenamefont{Ukai et~al.}(2010)\citenamefont{Ukai, Yoshikawa, Iwata,
  van Loock, and Furusawa}}]{ukai2010universal}
\bibinfo{author}{\bibfnamefont{R.}~\bibnamefont{Ukai {\it et al.}}},
  \bibinfo{journal}{Phys. Rev. A} \textbf{\bibinfo{volume}{81}},
  \bibinfo{pages}{32315} (\bibinfo{year}{2010}).

\bibitem{Raussendorf}
R.\ Raussendorf and H.~J.\ Briegel, Phys.\ Rev.\ Lett. {\bf 86}, 5188 (2001).

\bibitem{Menicucci}
N.~C.\ Menicucci {\it et al.},
Phys.\ Rev.\ Lett. {\bf 97}, 110501 (2006).


\bibitem[{\citenamefont{Yurke and Stoler}(1986)}]{Yurke1986}
\bibinfo{author}{\bibfnamefont{B.}~\bibnamefont{Yurke}} \bibnamefont{and}
  \bibinfo{author}{\bibfnamefont{D.}~\bibnamefont{Stoler}},
  \bibinfo{journal}{Phys. Rev. Lett.} \textbf{\bibinfo{volume}{57}},
  \bibinfo{pages}{13} (\bibinfo{year}{1986}).

\bibitem[{\citenamefont{Chuang and Yamamoto}(1995)}]{PhysRevA.52.3489}
\bibinfo{author}{\bibfnamefont{I.~L.} \bibnamefont{Chuang}} \bibnamefont{and}
  \bibinfo{author}{\bibfnamefont{Y.}~\bibnamefont{Yamamoto}},
  \bibinfo{journal}{Phys. Rev. A} \textbf{\bibinfo{volume}{52}},
  \bibinfo{pages}{3489} (\bibinfo{year}{1995}).

\bibitem{Klimov}
A.~B.\ Klimov, L.~L.\ S\'anches-Soto, and J.\ Delgado, Opt.\ Communic. {\bf 191}, 419 (2001).

\bibitem[{\citenamefont{Clark et~al.}(2003)\citenamefont{Clark, Lucarelli, and
  Tarn}}]{clark2003control}
\bibinfo{author}{\bibfnamefont{J.}~\bibnamefont{Clark}},
  \bibinfo{author}{\bibfnamefont{D.}~\bibnamefont{Lucarelli}},
  \bibnamefont{and} \bibinfo{author}{\bibfnamefont{T.}~\bibnamefont{Tarn}},
  \bibinfo{journal}{Int. J. Mod. Phys. B}
  \textbf{\bibinfo{volume}{17}}, \bibinfo{pages}{5397} (\bibinfo{year}{2003}).

\bibitem[{\citenamefont{Lloyd}(1995)}]{Lloyd1995}
\bibinfo{author}{\bibfnamefont{S.}~\bibnamefont{Lloyd}},
  \bibinfo{journal}{Phys. Rev. Lett.} \textbf{\bibinfo{volume}{75}},
  \bibinfo{pages}{346} (\bibinfo{year}{1995}).

\bibitem[{\citenamefont{Brown et~al.}(2010)\citenamefont{Brown, Munro, and
  Kendon}}]{brown2010using}
\bibinfo{author}{\bibfnamefont{K.}~\bibnamefont{Brown}},
  \bibinfo{author}{\bibfnamefont{W.}~\bibnamefont{Munro}}, \bibnamefont{and}
  \bibinfo{author}{\bibfnamefont{V.}~\bibnamefont{Kendon}},
  \bibinfo{journal}{Entropy} \textbf{\bibinfo{volume}{12}}, \bibinfo{pages}{2268} (\bibinfo{year}{2010}).

\bibitem[{\citenamefont{Suzuki}(1990)}]{suzuki1990fractal}
\bibinfo{author}{\bibfnamefont{M.}~\bibnamefont{Suzuki}},
  \bibinfo{journal}{Phys. Lett. A} \textbf{\bibinfo{volume}{146}},
  \bibinfo{pages}{319} (\bibinfo{year}{1990}).

\bibitem[{\citenamefont{McLachlan and Quispel}(2002)}]{McLachlan2002}
\bibinfo{author}{\bibfnamefont{R.~I.} \bibnamefont{McLachlan}}
  \bibnamefont{and} \bibinfo{author}{\bibfnamefont{R.~W.}
  \bibnamefont{Quispel}}, \bibinfo{journal}{Acta Numerica}
  \textbf{\bibinfo{volume}{11}}, \bibinfo{pages}{341} (\bibinfo{year}{2002}).

%


\bibitem{GKP}
D.\ Gottesman, A.\ Kitaev, and J.\ Preskill,
Phys.\ Rev.\ A {\bf 64}, 012310 (2001).

\bibitem{Ghose} S.\ Ghose and B.\ C.\ Sanders,
J.\ of Mod.\ Opt. {\bf 54}, 855 (2007).

\bibitem{Marek} P.\ Marek, R.\ Filip, and A.\ Furusawa,
arXiv:1105.4950.

\bibitem{MenicucciRalph}
N.~C.\ Menicucci, X.\ Ma, and T.~C.\ Ralph,
Phys.\ Rev.\ Lett. {\bf 104}, 250503 (2010).

\bibitem{Acin} A.\ J.\ Roncaglia {\it et al.},
Phys.\ Rev.\ A {\bf 83}, 062332 (2011).

\bibitem{Gu} M.\ Gu {\it et al.},
Phys.\ Rev.\ A {\bf 79}, 062318 (2009).

\bibitem{Eisaman} M.\ D.\ Eisaman {\it et al.},
Nature {\bf 438}, 837 (2005).

\bibitem{Chaneliere} T.\ Chaneliere {\it et al.},
Nature {\bf 438}, 833 (2005).

\bibitem{Polzik} B.\ Julsgaard {\it et al.},
Nature {\bf 432}, 482 (2004).

\bibitem{Fiurasek} J.\ Fiur\'{a}\u{s}ek,
Phys.\ Rev.\ A {\bf 68}, 022304 (2003).

\bibitem{Fleischhauer} Z.\ Kurucz and M.\ Fleischhauer,
Phys.\ Rev.\ A {\bf 78}, 023805 (2008).

\bibitem{UkaiPRL} R.\ Ukai {\it et al.},
Phys.\ Rev.\ Lett. {\bf 106}, 240504 (2011).

\bibitem[{\citenamefont{Yoshida}(1990)}]{yoshida1990construction}
\bibinfo{author}{\bibfnamefont{H.}~\bibnamefont{Yoshida}},
  \bibinfo{journal}{Phys. Lett. A} \textbf{\bibinfo{volume}{150}},
  \bibinfo{pages}{262} (\bibinfo{year}{1990}).
  
\bibitem[{\citenamefont{Reinsch}(2000)}]{reinsch2000simple}
\bibinfo{author}{\bibfnamefont{M.}~\bibnamefont{Reinsch}},
  \bibinfo{journal}{J. of Math. Phys.}
  \textbf{\bibinfo{volume}{41}}, \bibinfo{pages}{2434} (\bibinfo{year}{2000}).
  
\bibitem[{\citenamefont{Morgan}(2009)}]{Morgan2009}
\bibinfo{author}{\bibfnamefont{A.}~\bibnamefont{Morgan}},
  \emph{\bibinfo{title}{Solving polynomial systems using continuation for
  engineering and scientific problems}} (\bibinfo{publisher}{siam},
  \bibinfo{year}{2009}).

\bibitem[{\citenamefont{Suzuki}(1992)}]{Suzuki1992b}
\bibinfo{author}{\bibfnamefont{M.}~\bibnamefont{Suzuki}},
  \bibinfo{journal}{J. of the Phys. Soc. of Japan}
  \textbf{\bibinfo{volume}{61}}, \bibinfo{pages}{3015} (\bibinfo{year}{1992}).

\end{thebibliography}
\end{document}